# SELF-ORGANIZATION AND FRACTALITY IN A METABOLIC PROCESS OF THE KREBS CYCLE

*V.I. GRYTSAY*[a,*], *I.V. MUSATENKO*[b]

[a]*Bogolyubov Institute for Theoretical Physics, 14b, Metrolohichna Str., Kyiv 03680, Ukraine;*
*E-mail: vgrytsay@bitp.kiev.ua*
[b] *Taras Shevchenko National University of Kyiv, Faculty of Cybernetics, Department of Computational Mathematics, 64, Volodymyrska Str., Kyiv, Ukraine*
*E-mail: ivmusatenko@gmail.com*

*With the help of a mathematical model, the metabolic process of the Krebs cycle is studied. The autocatalytic processes resulting in both the formation of the self-organization in the Krebs cycle and the appearance of a cyclicity of its dynamics are determined. Some structural-functional connections creating the synchronism of an autoperiodic functioning at the transport in the respiratory chain and the oxidative phosphorylation are investigated. The conditions for breaking the synchronization of processes, increasing the multiplicity of a cyclicity, and for the appearance of chaotic modes are analyzed. The phase-parametric diagram of a cascade of bifurcations showing the transition to a chaotic mode by the Feigenbaum scenario is obtained. The fractal nature of the revealed cascade of bifurcations is demonstrated. The strange attractors formed as a result of the folding are obtained.*

*The results obtained give the idea of structural-functional connections, due to which the self-organization appears in the metabolic process running in a cell.*

*The constructed mathematical model can be applied to the study of the toxic and allergic effects of drugs and various substances on the metabolism of a cell.*

The search for general physical laws of self-organization in the Nature is one of the most important problems of natural sciences. As the principal problem, we mention the study of the nature of life, namely how the catalyzed enzymatic reactions create the internal space-time ordering of a cell. The development of mathematical models of such systems allows one to obtain a lot of important information.

Among various metabolic processes, the Krebs cycle is common for all cells [1]. The cycle involving tricarboxylic acids occupies the principal place, to which practically all metabolic paths lead. It is the final point of the catabolism of a "cell fuel," the central place of a cell respiration, and the source of molecules-predecessors, from which the aminoacids, carbohydrates, fat acids, and other compounds important for the functioning of cells are then synthesized. Its study will allow one to find the general regularities of the operation of a cell.

The functioning and the regulation of the Krebs cycle was studied experimentally and theoretically [2-10] within a restricted number of mathematical models. Many of the models considered only the stoichiometry and segments of the cycle стехонометрия и сегменты цикла. As the main difficulty, the deficit of experimental data should be mentioned.

It is difficult to study separately this cycle without the consideration of the functioning of a cell as a whole, since it is impossible to determine the internal parameters of a metabolic process without the knowledge of the external parameters of a medium, where a cell lives. Our studies are based on the mathematical model of the unstable growth of cells *Candida utilis* on ethanol, which was developed by Professor V.P. Gachok [11,12] in view of the experimental data published in [13]. Analogous problems, namely the modeling of a cell growth, were considered by J. Monod, V.S. Podgorskii, L.N. Drozdov-Tikhomirov, N.T. Rakhimova, G.Yu. Riznichenko, and others [14-18].

Our refined mathematical model involves the formation of carbon dioxide and considers its influence on the kinetics of a metabolic process.

In the present work, we will study the unstable modes of the cultivation of cells and will find the parameters and structural-functional connections of the metabolic process running in a cell that admit the appearance of oscillations. We will consider the intracellular autooscillations characteristic of the metabolic process on the level of redox reactions of the Krebs cycle. These autooscillations are related to the cyclicity of the process and characterize the self-organization inside a cell.

Analogous oscillatory modes are observed in the processes of photosynthesis and glycolysis, a variation in the concentration of calcium in a cell, oscillations in a heart muscle, and some biochemical processes [19-27].

We note that the description of the metabolic process within the presented model is based not on the use of conditional reagents, as was made earlier in other models, but involves the real intracellular components, which will allow us to find an actual mechanism of self-organization.

## MATHEMATICAL MODEL

The general scheme of the process is presented in Fig. 1. According to it with regard for the mass balance, we have constructed the mathematical model given by Eqs. (1) - (19).

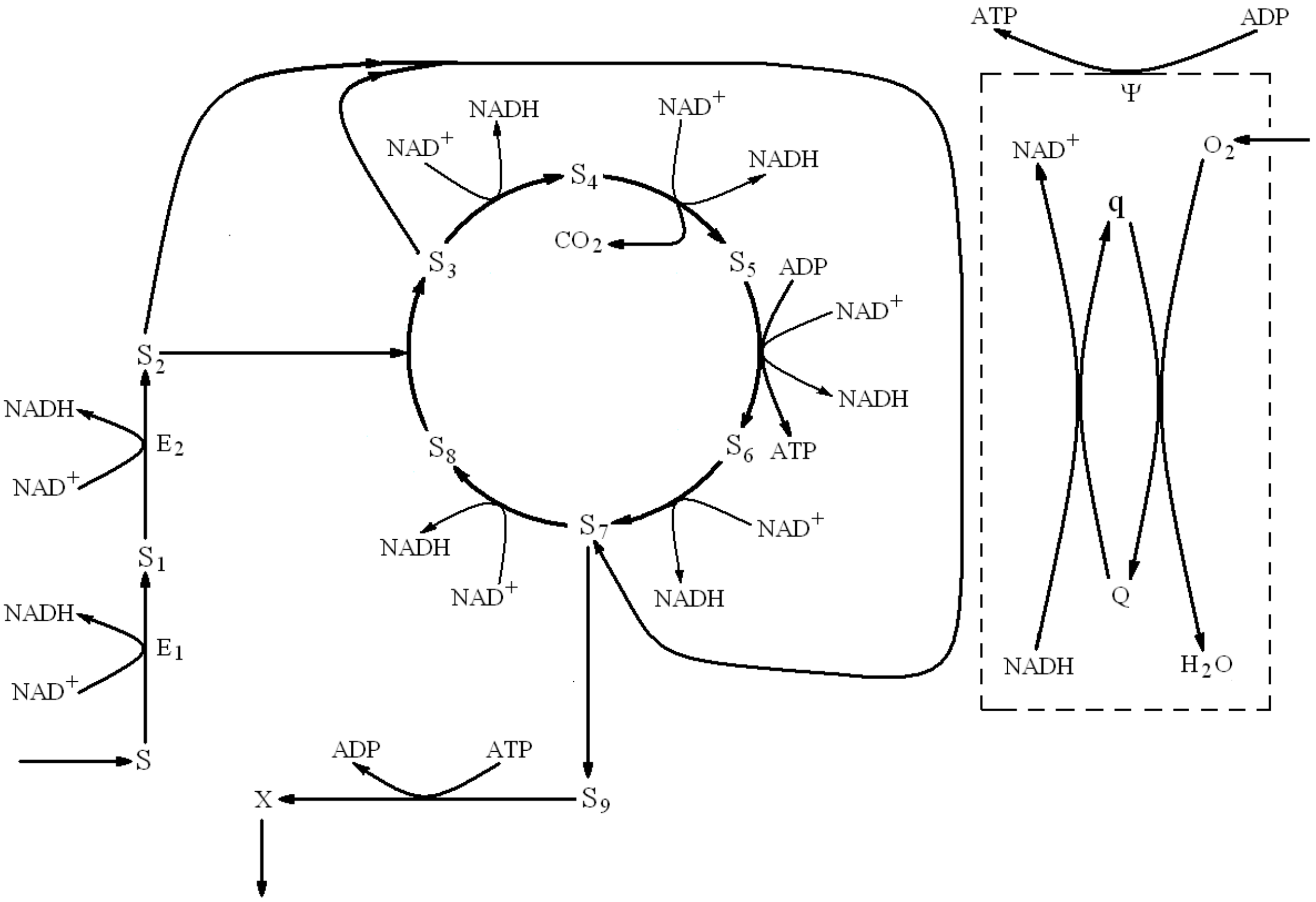

Fig. 1. General scheme of the metabolic process of growth of cells *Candida utilis* on ethanol [12].

$$\frac{dS}{dt} = S_0 \frac{K}{K + S + \gamma\psi} - k_1 V(E_1) \frac{N}{K_1 + N} V(S) - \alpha_1 S, \tag{1}$$

$$\frac{dS_1}{dt} = k_1 V(E_1) \frac{N}{K_1 + N} V(S) - k_2 V(E_2) \frac{N}{K_1 + N} V(S_1), \tag{2}$$

$$\frac{dS_2}{dt} = k_2 V(E_2) \frac{N}{K_1 + N} V(S_1) - k_3 V(S_2^2) V(S_3) - k_4 V(S_2) V(S_8), \tag{3}$$

$$\frac{dS_3}{dt} = k_4 V(S_2) V(S_8) - k_5 V(N^2) V(S_3^2) - k_3 V(S_2^2) V(S_3), \tag{4}$$

$$\frac{dS_4}{dt} = k_5 V(N^2) V(S_3^2) - k_7 V(N) V(S_4) - k_8 V(N) V(S_4), \tag{5}$$

$$\frac{dS_5}{dt} = k_7 V(N) V(S_4) - 2k_9 V(L_1 - T) V(S_5), \tag{6}$$

$$\frac{dS_6}{dt} = 2k_9 V(L_1 - T) V(S_5) - k_{10} V(N) \frac{S_6^2}{S_6^2 + 1 + M_1 S_8}, \tag{7}$$

$$\frac{dS_7}{dt} = k_{10} V(N) \frac{S_6^2}{S_6^2 + 1 + M_1 S_8} - k_{11} V(N) V(S_7) - k_{12} \frac{S_7^2}{S_7^2 + 1 + M_2 S_9} V(\psi^2) + k_3 V(S_2^2) V(S_3), \tag{8}$$

$$\frac{dS_8}{dt} = k_{11} V(N) V(S_7) - k_4 V(S_2) V(S_8) + k_6 V(T^2) \frac{S^2}{S^2 + \beta_1} \cdot \frac{N_1}{N_1 + (S_5 + S_7)^2}, \tag{9}$$

$$\frac{dS_9}{dt} = k_{12} \frac{S_7^2}{S_7^2 + 1 + M_2 S_9} V(\psi^2) - k_{14} \frac{XTS_9}{(\mu_1 + T)[(\mu_2 + S_9 + X + M_3(1 + \mu_3 \psi)]S}, \tag{10}$$

$$\frac{dX}{dt} = k_{14} \frac{XTS_9}{(\mu_1 + T)[(\mu_2 + S_9 + X + M_3(1 + \mu_3 \psi)]S} - \alpha_2 X, \tag{11}$$

$$\frac{dQ}{dt} = -k_{15} V(Q) V(L_2 - N) + 4k_{16} V(L_3 - Q) V(O_2) \frac{1}{1 + \gamma_1 \psi^2}, \tag{12}$$

$$\frac{dO_2}{dt} = O_{2_0} \frac{K_2}{K_2 + O_2} - k_{16}(L_3 - Q) V(O_2) \frac{1}{1 + \gamma_1 \psi} - k_8 V(N) V(S_4) - \alpha_3 O_2, \tag{13}$$

$$\frac{dN}{dt} = -k_7 V(N) V(S_4) - k_{10} V(N) \frac{S_6^2}{S_6^2 + 1 + M_1 S_8} - k_{11} V(N) V(S_7) - k_5 V(N^2) V(S_3^2) +$$

$$+ k_{15} V(Q) V(L_2 - N) - k_2 V(E_2) \frac{N}{K_1 + N} V(S_1) - k_1 V(E_1) \frac{N}{K_1 + N} V(S), \tag{14}$$

$$\frac{dT}{dt} = k_{17} V(L_1 - T) V(\psi^2) + k_9 V(L - T) V(S_3) - \alpha_4 T -$$

$$- k_{18} k_6 V(T^2) \frac{S^2}{S^2 + \beta_1} \cdot \frac{N_1}{N_1 + (S_5 + S_7)^2} - k_{19} k_{14} \frac{XTS_9}{(\mu_1 + T)[\mu_2 + S_9 + X + M_3(1 + \mu_3 \psi) S]}, \tag{15}$$

$$\frac{d\psi}{dt} = 4k_{15} V(Q) V(L_2 - N) + 4k_{17} V((L_1 - T) V(\psi^2) - 2k_{12} \frac{S_7^2}{S_7^2 + 1 + M_2 S_9} V(\psi^2) - \alpha \psi, \tag{16}$$

$$\frac{dE_1}{dt} = E_{1_0}\frac{S^2}{\beta_2 + S^2}\frac{N_2}{N_2 + S_1} - n_1 V(E_1)\frac{N}{K_1 + N}V(S) - \alpha_5 E_1, \tag{17}$$

$$\frac{dE_2}{dt} = E_{2_0}\frac{S_1^2}{\beta_3 + S_1^2}\frac{N_3}{N_3 + S_2} - n_2 V(E_2)\frac{N}{K_1 + N}V(S_1) - \alpha_6 E_2, \tag{18}$$

$$\frac{dC}{dt} = k_8 V(N)V(S_4) - \alpha_7 C. \tag{19}$$

where $V(X) = X/(1+X)$ is the function that describes the adsorption of the enzyme in the region of a local coupling. The variables of the system are dimensionless.

The internal parameters of the system are as follows:

$k_1 = 0.3;\ k_2 = 0.3;\ k_3 = 0.2;\ k_4 = 0.6;\ k_5 = 0.16;\ k_6 = 0.7;\ k_7 = 0.08;\ k_8 = 0.022;$

$k_9 = 0.1;\ k_{10} = 0.08;\ k_{11} = 0.08;\ k_{12} = 0.1;\ k_{14} = 0.7;\ k_{15} = 0.27;\ k_{16} = 0.18;$

$k_{17} = 0.14;\ k_{18} = 1;\ k_{19} = 10;\ n_1 = 0.07;\ n_2 = 0.07;\ L = 2;\ L_1 = 2;\ L_2 = 2.5;\ L_3 = 2;$

$K = 2.5;\ K_1 = 0.35;\ K_2 = 2;\ M_1 = 1;\ M_2 = 0.35;\ M_3 = 1;\ N_1 = 0.6;\ N_2 = 0.03;$

$N_3 = 0.01;\ \mu_1 = 1.37;\ \mu_2 = 0.3;\ \mu_3 = 0.01;\ \gamma = 0.7;\ \gamma_1 = 0.7;\ \beta_1 = 0.5;\ \beta_2 = 0.4;$

$\beta_3 = 0.4;\ E_{1_0} = 2;\ E_{2_0} = 2.$

The external parameters determining the flow-type conditions are chosen as

$S_0 = 0.05055;\ O_{2_0} = 0.06;\ \alpha = 0.002;\ \alpha_1 = 0.02;\ \alpha_2 = 0.004;\ \alpha_3 = 0.01;$

$\alpha_4 = 0.01;\ \alpha_5 = 0.01;\ \alpha_6 = 0.01;\ \alpha_7 = 0.0001.$

The model covers the processes of substrate-enzymatic oxidation of ethanol to acetate, cycle involving tri- and dicarboxylic acids, glyoxylate cycle, and respiratory chain. Model (1)-(19) is improved as compared with the model used in [11-12], since it involves the formation of $CO_2$ in of the Krebs cycle, which affects the running of the metabolic process. Some parameters of our model are taken from [11-12].

The incoming ethanol $S$ is oxidized by the alcohol dehydrogenase enzyme $E_1$ to acetaldehyde $S_1$ (1) and then by the acetal dehydrogenase enzyme $E_2$ to acetate $S_2$ (2), (3). The formed acetate can participate in the cell metabolism and can be exchanged with the environment. The model accounts for this situation by the change of acetate by acetyl-$CoA$. On the first stage of the Krebs cycle due to the citrate synthase reaction, acetyl-$CoA$ jointly with oxalacetate $S_8$ formed in the Krebs cycle produce citrate $S_3$ (4). Then substances $S_4$ - $S_8$ are created successively on stages (5)- (9). In the model, the Krebs cycle is represented by only those substrates that participate in the reduction of $NADH$ and the phosphorylation $ADT \to ATP$. Acetyl-$CoA$ passes along the chain to malate represented in the model as intramitochondrial $S_7$ (8) and cytosolic $S_9$ (10) ones. Malate can be also synthesized in another way related to the activity of two enzymes: isocitrate lyase and malate synthetase. The former catalyzes the splitting of isocitrate to succinate, and the latter catalyzes the condensation of acetyl-$CoA$ with glyoxylate and the formation of malate. This glyoxylate-linked way is shown in Fig. 1 as an enzymatic reaction with the consumption of $S_2$ and $S_3$ and the formation of $S_7$. The parameter $k_3$ controls the activity of the активность glyoxylate-linked way (3), (4), (8). The yield of $S_7$ into cytosol is controlled by its concentration, which can increase due to $S_9$, by causing the inhibition of its transport with the participation of protons of mitochondrial membrane.

The formed malate $S_9$ is used by a cell for its growth, namely for the biosynthesis of protein $X$ (11). The energy consumption of the given process is supported by the process $ATP \to ADP$. The presence of ethanol in the external solution causes the "ageing" of external membranes of cells, which leads to the inhibition of this process. The inhibition of the process also happens due to the enhanced level of the kinetic membrane potential $\psi$. The parameter $\mu_0$ is related to the lysis and the washout of cells.

In the model, the respiratory chain of a cell is represented in two forms: oxidized, $Q$, (12) and reduced, $q$, ones. They obey the integral of motion $Q(t) + q(t) = L_3$.

A change of the concentration of oxygen in the respiratory chain is determined by Eq. (13).

The activity of the respiratory chain is affected by the level of $NADH$ (14). Its high concentration leads to the enhanced endogenic respiration in the reducing process in the respiratory chain (parameter $k_{15}$). The accumulation of $NADH$ occurs as a result of the reduction of $NAD^+$ at the transformation of ethanol and in the Krebs cycle. These variables obey the integral of motion $NAD^+(t) + NADH(t) = L_2$.

In the respiratory chain and the Krebs cycle, the substrate-linked phosphorylation of $ADP$ with the formation of $ATP$ (15) is also realized. The energy consumption due to the process $ATP \to ADP$ induces the biosynthesis of components of the Krebs cycle (parameter $k_{18}$) and the growth of cells on the substrate (parameter $k_{19}$). For these variables, the integral of motion $ATP(t) + ADP(t) = L_1$ holds. Thus, the level of $ATP$ produced in the redox processes in the respiratory chain $ADP \to ATP$ determines the intensity of the Krebs cycle and the biosynthesis of protein.

In the respiratory chain, the kinetic membrane potential $\psi$ (16) is created under the running of reducing processes $Q \rightarrow q$. It is consumed at the substrate-linked phosphorylation $ADP \rightarrow ATP$ in the respiratory chain and the Krebs cycle. Its enhanced level inhibits the biosynthesis of protein and process of reduction of the respiratory chain.

Equations (17) and (18) describe the activity of enzymes $E_1$ and $E_2$, respectively. We consider their biosynthesis ($E_{1_0}$ and $E_{2_0}$), the inactivation in the course of the enzymatic reaction ($n_1$ and $n_2$), and all possible irreversible inactivations ($\alpha_5$ and $\alpha_6$).

Equation (19) is related to the formation of carbon dioxide. Its removal from the solution into the environment ($\alpha_C$) is taken into account. Carbon dioxide is produced in the Krebs cycle (5). In addition, it squeezes out oxygen from the solution (13), by decreasing the activity of the respiratory chain.

The study of system (1)-(19) was carried out with the help of the Runge-Kutta-Merson method. The analysis of modes was peformed when the system approaches the asymptotic trajectory of an attractor. To this end, the calculations started after *1,000,000* preliminary steps. The solution accuracy was set to be $10^{-12}$.

In our studies, we used the classical tools of nonlinear dynamics [28,29] and some methods of simulation of biochemical processes developed by the authors in [30-35].

## RESULTS OF STUDIES

According to the constructed mathematical model (1)-(19), it is possible to separate two autocatalytic processes creating the space-time self-organization of the mutually connected metabolic processes of the Krebs cycle and the respiratory chain.

The continuous operation of the Krebs cycle requires the permanent supply of acetyl-$CoA$ into the cycle. This substance is formed in the catabolism of ethanol, being the "cell fuel." Then acetyl-$CoA$ is condensed with oxalocitrate and forms citric acid. After the dehydrogenation and two decarboxylations, citric acid is gradually transformed along a circle ($S_3 \rightarrow S_8$) to $CO_2$ and $H_2O$. In this case, the regeneration of oxalocitrate occurs at the end. On each stage of this cycle, there occurs the reduction of co-enzyme $NAD^+$ to $NADH$. The reverse process of oxidation $NADH \rightarrow NAD^+$ is realized in the system of carriers of electrons in the respiratory chain (1-st autocatalytic cycle).

The energy released as a result of the oxidation of acetyl-$CoA$ is concentrated to a significant degree in macroergic phosphate bonds of $ATP$. This happens at the transfer of electrons from the Krebs cycle to the respiratory chain, where the oxidative phosphorylation $ADT \rightarrow ATP$ occurs.

The formed $ATP$ is consumed for the growth of cells and in the Krebs cycle itself $ATP \to ADP$ (2-nd autocatalytic cycle).

The preservation of the functioning of the Krebs cycle depends of the key parameters of the system relating to the inverse paths of the reduction of oxalocitrate. They are the parameters $k_{15}$ and $k_6$. The former determines the rate of oxidation, $NADH \to NAD^+$, in the respiratory chain and depends on its parameters: $Q(t) + q(t) = L_3$. The rate of the whole autocatalytic process $NADH \leftrightarrow NAD^+$ is determined by the rate of the redox process in the respiratory chain itself $Q \leftrightarrow q$.

The parameter $k_6$ determines the influence of the concentration of $ADP$ on the rate of the metabolic process of the Krebs cycle.

Thus, the metabolic process in the Krebs cycle is characterized by the appearance of autooscillations, whose frequency and stability depend on the synchronization of the 1-st and 2-nd autocatalytic cycles. Let us consider the kinetics of autooscillations in the metabolic process of the Krebs cycle, which arise in its cyclic mode. We will study the stability of the Krebs cycle on the whole.

During one turnover of a cycle, one molecule of acetyl-$CoA$ is completely oxidized to malate, and a new molecule of acetyl-$CoA$ is formed at the input. In such a way, the continuous process of operation of the Krebs cycle is running and manifests the autooscillatory character. In Fig. 2, we show the kinetics of autooscillations of components of the cycle. At $k_{15}$=0.23, a 1-fold periodic cycle appears in the system. The oscillations of components occur synchronously with phase shifts corresponding to the space-time positions of these components in the cycle. The oscillations at the same frequency will occur in all components of the given metabolic process.

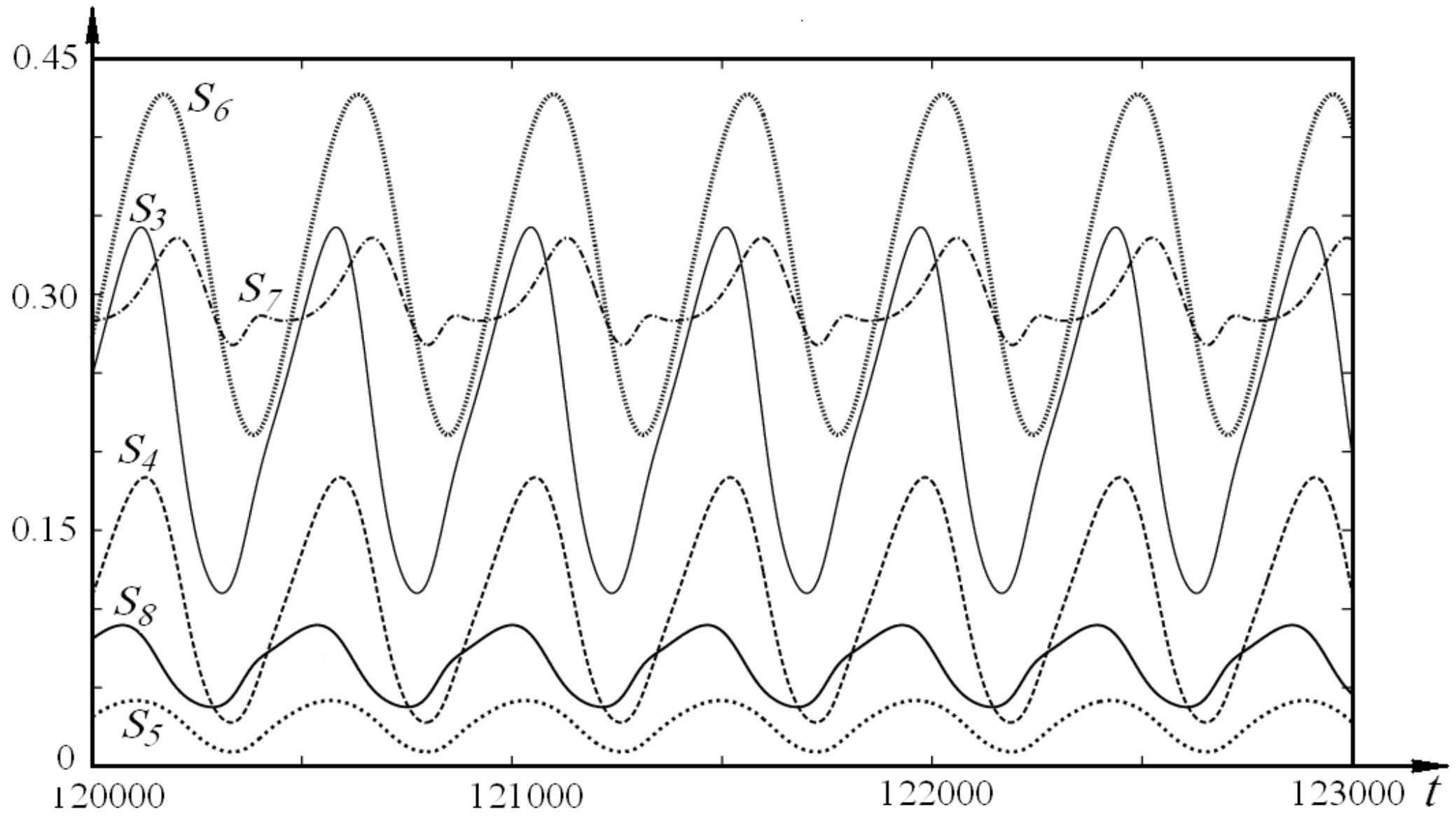

Fig. 2. Kinetics of autooscillations of components of the Krebs cycle $S_3 - S_8$, at $k_{15}$=0.23.

In Fig. 3,a, we show a phase-parametric diagram presenting the dependence of the multiplicity of autooscillations of the input substrate, ethanol ($S$), on the parameter $k_{15}$. To construct the phase-parametric diagram, we used the method of cutting. In the phase space of trajectories of the system, we placed a cutting plane at $S_2 = 0.8$. If the trajectory crosses this plane in a certain direction, we indicate the value of chosen variable ($S$, in this case) on the phase-parametric diagram. It is seen from the phase-parametric diagram that, at $k_{15}^{j} = 0.2328$, the doubling of the period of oscillations arises. At $k_{15}^{j+1} = 0.2483$, we observe the repeated period doubling. Then, at $k_{15}^{j+2} = 0.2516$, the period of autooscillations is doubled one more.

Let us separate a small part of the diagram with $k_{15} \in (0.25, 0.254)$ (Fig. 3,a) and magnify it (Fig. 3,b). This part of the phase-parametric diagram is identical to the whole diagram in Fig. 3,a. This indicates that, after the ($j+2$)-th bifurcation, the next bifurcation arises, and so on to ($j+n$). In other words, we have a cascade of bifurcations. At the following decrease in the scale of the diagram, the pattern will repeat up to the critical value of parameter $k_{15}$, after which we observe the appearance of a deterministic xaoc. For the derived sequence of bifurcations, we have the relation

$$\lim_{t\to\infty} \frac{k_{15}^{j+1} - k_{15}^{j}}{k_{15}^{j+2} - k_{15}^{j+1}} \approx 4.696969\ldots.$$

This number is very close to the Feigenbaum universal constant $\delta$ = 4.669211660910…, which characterizes the infinite cascade of bifurcations at the transition to a deterministic xaocy. Thus, the part under consideration is characterized by the doubling of the period of autooscillations by the Feigenbaum scenario [36-38] at the increase in the dissipation coefficient $\alpha$. This means that, for the given unstable modes of the physical system, any arising fluctuation can lead to a chaotic cyclic mode.

At $k_{15} = 0.2541$ and $k_{15} = 0.2603$ (Fig. 3,a), the windows of periodicity appear. The deterministic chaos is destroyed, and periodic and quasiperiodic modes are established. Outside these windows, chaotic modes hold. The identical windows of periodicity are observed also on less scales of the diagram (Fig. 3,b). In other words, the phase-parametric diagrams on a small scale (Fig. 3,b) and on a large one (Fig. 3,a) are analogous. This fact indicates the fractal nature of the obtained cascade of bifurcations.

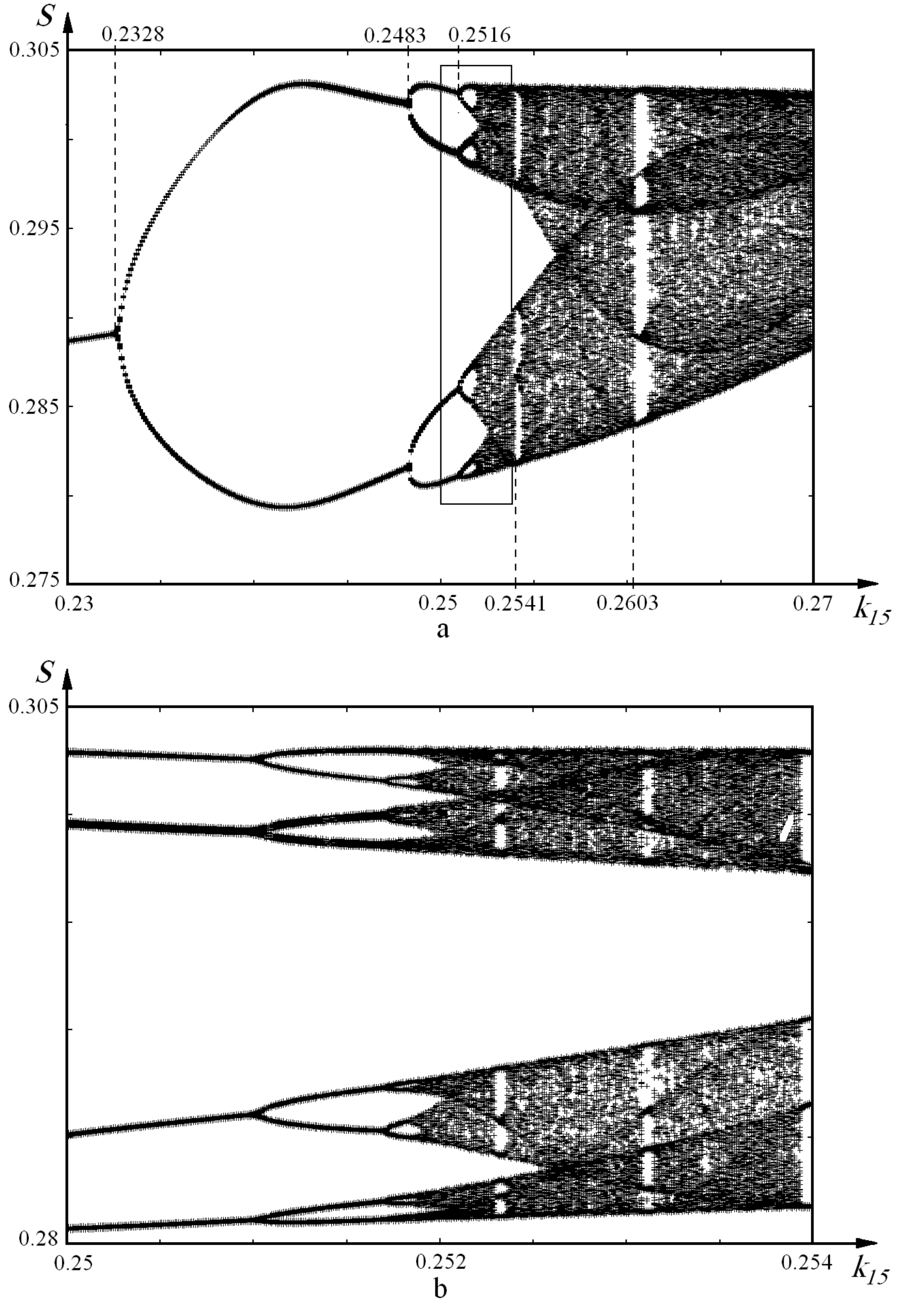

Fig. 3. Phase-parametric diagram of the system: (a) - $k_{15} \in (0.23, 0.27)$; (b) - $k_{15} \in (0.25, 0.254)$.

As examples of the successive period doubling for the autoperiodic modes of the system by the the Feigenbaum scenario, w present the projections of the phase portraits of appropriate regular attractors in Fig. 4,a-d. In Fig. 4,e, we show a regular attractor of the 6-fold quasiperiodic mode $\approx 6 \cdot 2^0$ arising in the window of periodicity at $k_{15} = 0.2543$ (Fig. 3,a). Figure 4,f represents strange attractor $2 \cdot 2^x$ ($k_{15} = 0.267$).

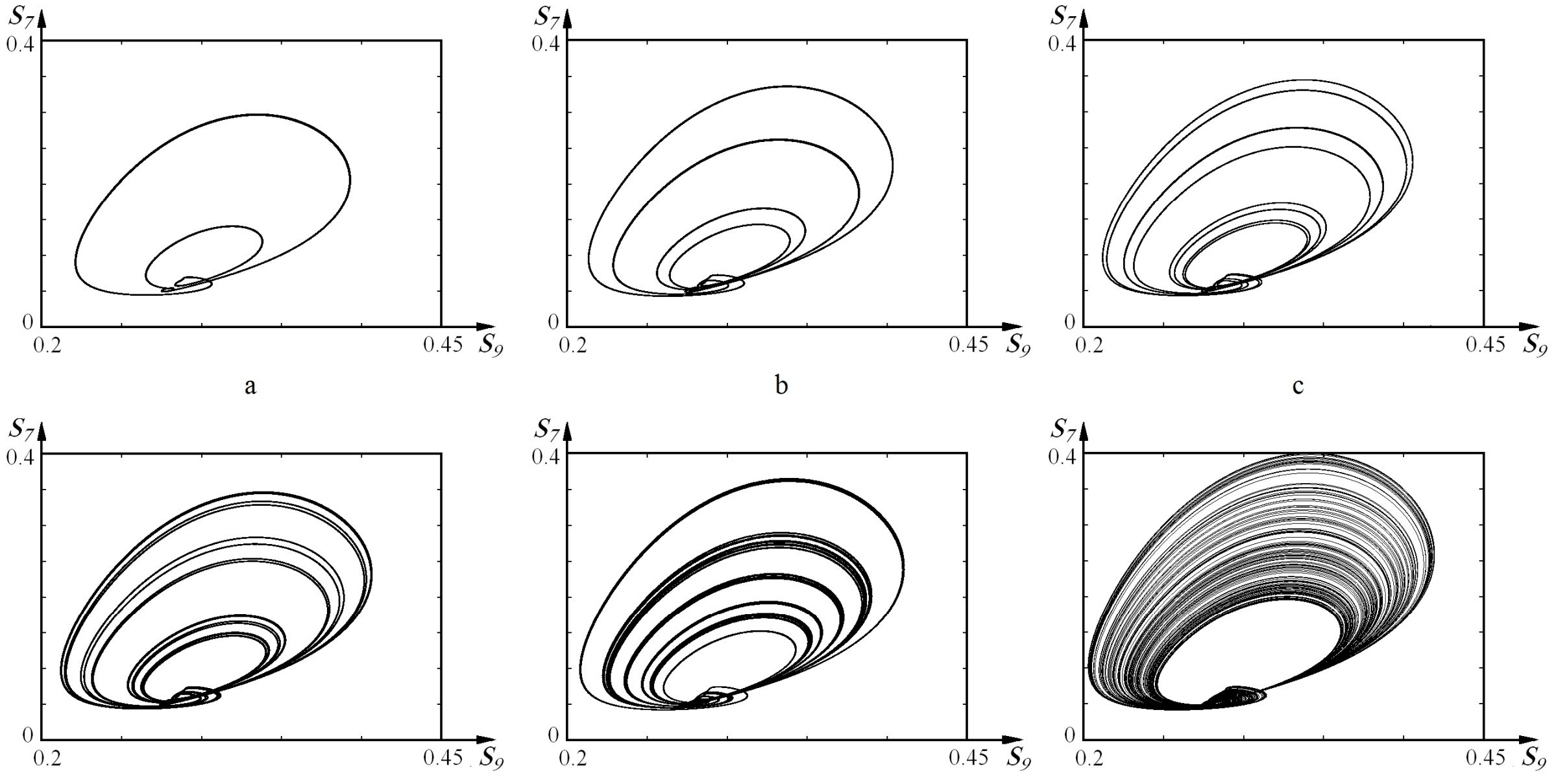

Fig. 4. Projections of the phase portraits of attractors of the system: (a) - $2 \cdot 2^0$ ($k_{15} = 0.2482$); (b) - $4 \cdot 2^0$ ($k_{15} = 0.251$); (c) - $8 \cdot 2^0$ ($k_{15} = 0.2517$); (d) - $16 \cdot 2^0$ ($k_{15} = 0.2518$); (e) - $\approx 6 \cdot 2^0$ ($k_{15} = 0.2543$); (f) - $2 \cdot 2^x$ ($k_{15} = 0.267$).

The kinetics of changes in some components of the mode with strange attractor $2 \cdot 2^x$ ($k_{15} = 0.27$) is shown in Fig. 5. As compared with Fig. 2, the synchronism of oscillations of citrate $S_3$ and oxalacetate $S_8$ at the input and the output of the Krebs cycle is violated. The cycle stops to be strictly periodic (Fig. 2). The periodicity and the amplitudes change and become chaotic (Fig. 5). This occurs due to the violation of the synchronism of metabolic processes in the system of transport of electrons ($NADH \leftrightarrow NAD^+$) of the redox process in the respiratory chain ($Q \leftrightarrow q$) and the substrate-linked phosphorylation $ADP \rightarrow ATP$.

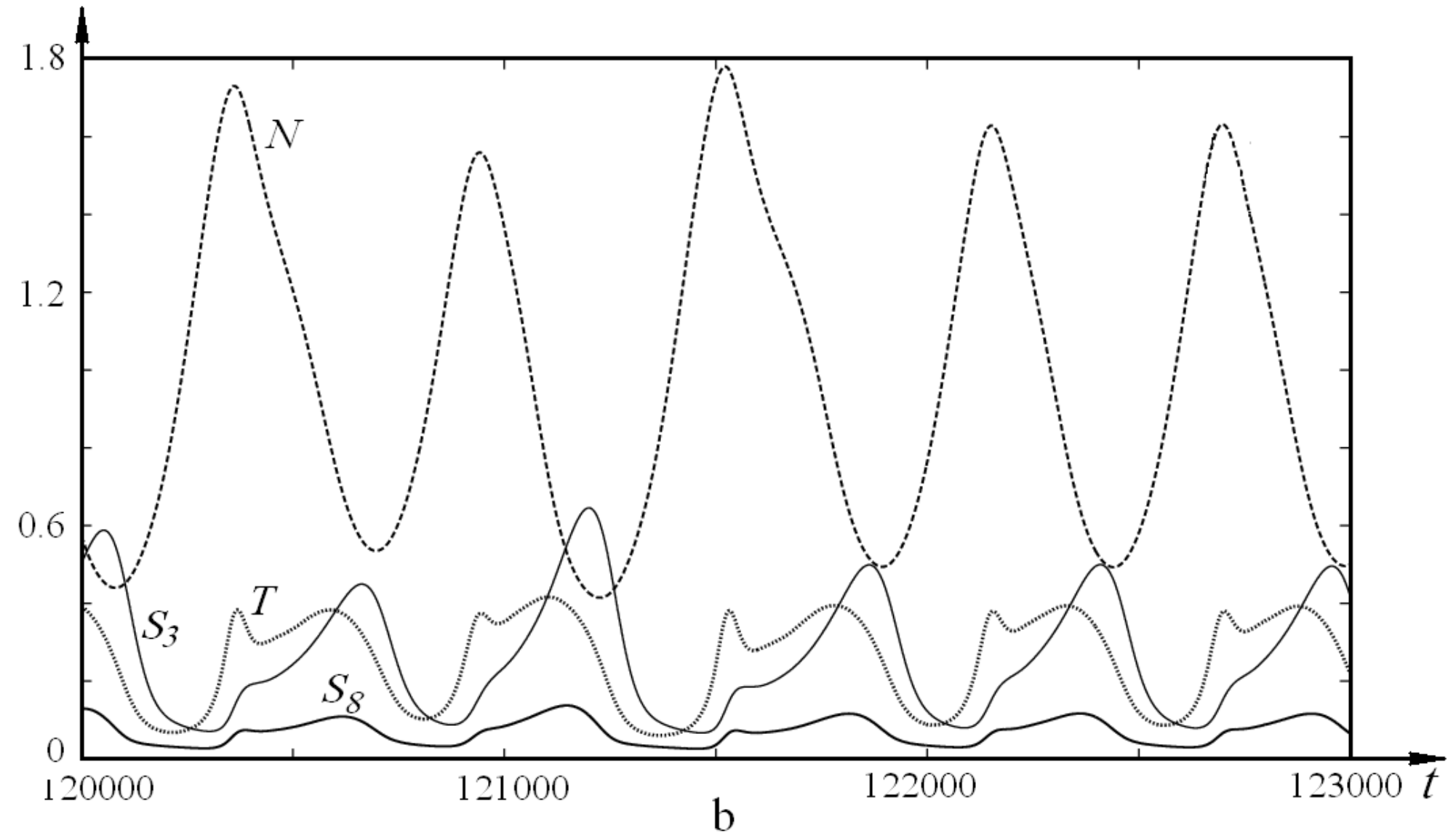

Fig. 5. Kinetic curves of components of the Krebs cycle $S_3$, $S_8$, $N$, and $T$ in the mode of strange attractor $2 \cdot 2^x$ at $k_{15}$=0.27 and $S_8^0 = 0.001$.

In Fig. 6,a,b, we show the projections of the phase portraits of the given strange attractor in the three-dimensional coordinate systems. The strange attractors obtained in this case are formed as a result of the folding. An element of the phase volume of such an attractor is stretched in some directions and shrinks in other ones, by conserving its stability. Therefore, we observe the mixing of trajectories in narrow places of the phase space of the system, i.e., a deterministic chaos arises.

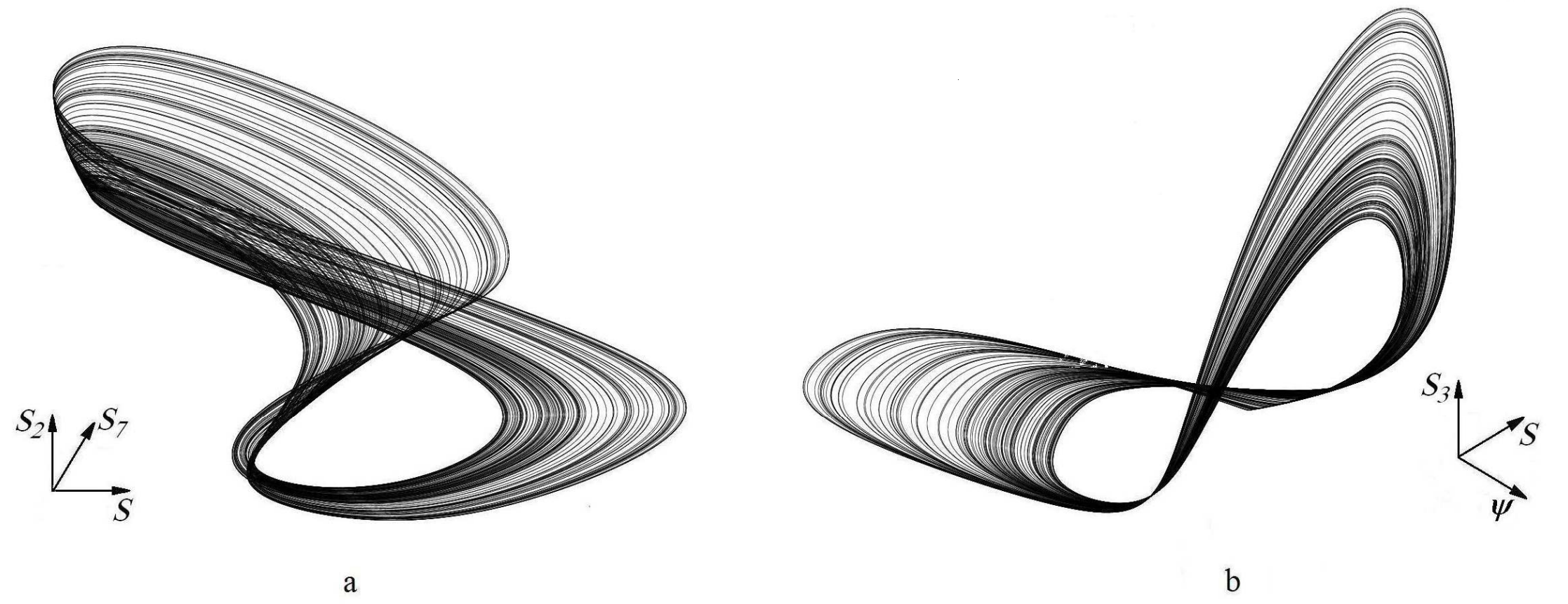

Fig. 6. Projections of the phase portraits of strange attractor $2 \cdot 2^x$ at $k_{15}$=0.27 in the three-dimensional coordinate systems: $(S, S_7, S_2)$ (a) and $(\psi, S, S_3)$ (b).

The whole Krebs cycle consists of 8 successive reactions. In its nodal points, there occurs the transfer of electrons through $NADH$ onto the respiratory chain. An increase in the parameter $k_{15}$ causes an increase in the rate of the process $NADH \leftrightarrow NAD^+$ on various parts of the cycle. This causes a change in the levels of relevant metabolites on the given part. Their recurrence period in the Krebs cycle is doubled, by preserving a cyclicity by the Feigenbaum scenario. In addition, the rate of substrate-linked phosphorylation $ADP \to ATP$ in the respiratory chain is also increased. The arisen additional peaks on the plot of $ATP$ correspond to those on the plot of $NAD^+$ (Fig. 5). Since $ADP$ takes participation in the Krebs cycle, this causes nonlinear variations of the periodic kinetics of metabolites. The cyclicity of restoration of the level of oxalacetate $S_8$ is violated, which affects the subsequent cycles. In this case, a chaotic mode arises. Thus, there occurs the permanent adaptation of the metabolic process of a cell to variations in the environment. This leads to the adaptation of the whole metabolic process of a cell and to its self-organization for the conservation of homeostasis.

The results obtained allow one to estimate the stability of the metabolic process of a cell under the action of external factors and to forecast the consequence of this action on the vital activity of alive organisms. The developed mathematical model can be useful for the creation of new drugs and in the study of the toxicity of various substances.

*The work is supported by the project N 0113U001093 of the National Academy of Sciences of Ukraine.*

## САМООРГАНІЗАЦІЯ ТА ФРАКТАЛЬНІСТЬ В МЕТАБОЛІЧНОМУ ПРОЦЕСІ ЦИКЛУ КРЕБСА

*В.Й. Грицай*[a,*], *І.В. Мусатенко*[b]

[a]*Інститут теоретичної фізики ім. М.М. Боголюбова, 14б, вул. Метрологічна, Київ 03680, Україна*

*E-mail: vgrytsay@bitp.kiev.ua*

[b] *Київський Національний університет ім. Тараса Шевченко, Факультет кібернетики, Кафедра обчислювальної математики, 64, вул. Володимирська, Київ, Україна*

*E-mail: ivmusatenko@gmail.com*

*В роботі за допомогою математичної моделі досліджується метаболічний процес циклу Кребса. Знайдено автокаталітичні процеси, внаслідок яких утворюється самоорганізація в циклі Кребса і виникає циклічність в його динаміці. Досліджено структурно-функціональні зв'язки, які створюють синхронність автоперіодичного функціонування при переносі електронів в дихальному ланцюзі та при окислювальному фосфорилюванні. Досліджено*

*умови порушення синхронізації процесів, збільшення кратності циклічності та виникнення хаотичних режимів.Отримана фазопараметрична діаграма каскаду біфуркацій, що відображує перехід до хаотичного режиму відповідно сценарію Фейгенбаума. Показано фрактальну природу знайденого каскаду біфуркацій. Знайдено дивні атрактори, що утворюються завдяки формуванню складки.*

*Отримані результати дають представлення про структурно-функціональні зв'язки, завдяки яким виникає самоорганізація в метаболічному процесі клітини.*

*Побудована математична модель може бути використана для дослідження токсичного та алергічного впливу ліків та різних речовин на метаболізм клітини.*

## САМООРГАНИЗАЦИЯ И ФРАКТАЛЬНОСТЬ В МЕТАБОЛИЧЕСКОМ ПРОЦЕССЕ ЦИКЛА КРЕБСА

*В.И. Грицай[a,*], И.В. Мусатенко[b]*

*[a]Институт теоретической физики им. Н.Н. Боголюбова, 14б, ул. Метрологическая, Киев 03680, Украина*

*E-mail: vgrytsay@bitp.kiev.ua*

*[b] Киевский Национальный университет им. Тараса Шевченко, Факультет кибернетики, Кафедра вычислительной математики, 64, ул. Владимирская, Киев, Украина*

*E-mail: ivmusatenko@gmail.com*

*В работе при помощи математической модели исследуется метаболический процесс цикла Кребса. Найдены автокаталитические процессы, вследствие которых образуется самоорганизация в цикле Кребса и возникает цикличность в его динамике. Исследованы структурно-функциональные связи, создающие синхронность автопериодического функционирования при переносе электронов в дыхательной цепи и при окислительном фосфорилировании. Исследованы условия нарушения синхронизации процессов, увеличения кратности цикличности и возникновения хаотических режимов. Получена фазопараметрическая диаграмма каскада бифуркаций, отражающая переход к хаотическому режиму согласно сценарию Фейгенбаума. Показана фрактальная природа*

*найденного каскада бифуркаций. Найдены странные аттракторы, образуемые в результате образования складки.*

*Полученные результаты дают представление о структурно-функциональных связях, благодаря которым возникает самоорганизация в метаболическом процессе клетки.*

*Построенная математическая модель может быть использована для исследований токсического и аллергического влияния лекарств и различных веществ на метаболизм клетки.*

*Ключевые слова: цикл Кребса, метаболический процесс, самоорганизация,фрактальность, странный аттрактор, сценарий Фейгенбаума.*